# Learning with multiple representations:
# an example of a revision lesson in mechanics

[1]Darren Wong, [2]Sng Peng Poo, [2]Ng Eng Hock and [3]Wee Loo Kang

[1]Natural Sciences and Science Education, National Institute of Education, Singapore.
[2]Anderson Junior College, Singapore.
[3]Educational Technology Division, Ministry of Education, Singapore.
darren.wong@nie.edu.sg, sng_peng_poo@moe.edu.sg, ng_eng_hock@moe.edu.sg, wee_loo_kang@moe.gov.sg

**Abstract**

We describe an example of learning with multiple representations in an A-level revision lesson on mechanics. The context of the problem involved the motion of a ball thrown upwards in air and studying how the associated physical quantities changed along its flight. Different groups of students were assigned to look at different aspects of the ball's motion using various representations: vector diagrams, free-body diagrams, verbal description, equations and graphs, drawn against time as well as against displacement. Overall, feedback from students about the lesson was positive. We further discuss the benefits of using computer simulation to support and extend student learning.

**Introduction**

Physics often involves the modelling of real world physical phenomena using external representations that range from concrete to abstract forms: pictures, diagrams, words, graphs and equations [1]. Indeed, new representational tools can be developed to scaffold student learning from the more concrete physical situation to the more abstract, but generalisable forms of representation. For example, Hinrichs [2] described how the system schema can serve as a conceptual bridge between the pictorial representation and the free-body diagram, to help students better understand the application of Newton's third law[1].

The general consensus from research indicates that multiple representations (MRs) play an important role in student learning by facilitating the acquisition of knowledge and guiding in problem solving [3]. A learner equipped with thinking in more than one representation is able to reason more flexibly when learning new material or solving a problem. Moreover, certain combinations of MRs can help learning when the interpretation of an unfamiliar or more abstract representation is constrained by a more familiar one. For example, concrete animations are often employed in simulations alongside complex representations such as graphs. MRs can also support the construction of deeper understanding when learners integrate information from the various representations to achieve insights that otherwise would be difficult to achieve with only a single representation. Ainsworth [4] has succinctly categorised the three key functions that MRs serve in learning as having *complementary* roles, *constraining* interpretation and *constructing* deeper understanding.

Many educators have adopted MRs as an instructional strategy and its use is often associated with the Modelling Instruction approach to learning physics [5], which emphasises the development of a sound conceptual understanding through diagrammatic and graphical representations before moving on to an algebraic treatment of problem solving. Others like Angell [6] developed a modelling teaching approach to upper secondary physics that emphasises scientific reasoning based on empirical data and using MRs of physical phenomenon as a teaching framework. Bryan & Fennel [7] developed an instructional sequence on wave motion that systematically guides students to the mathematical model behind the physical phenomenon, through a series of modelling activities.

---

[1] This is achieved by identifying and labeling all objects of interest and the different types of interactions between objects.

However, it has been argued that in order to reap the full benefits of MRs as a learning tool, learners need to be engaged in meaningful tasks associated with their use [4]. Learners must learn the *format* of the representation and understand how the representation *relates* to the specific topic it is representing (see Table 1 for typical visual representations that are associated with specific topics). Additionally, learners have to be able to *interpret* given representations, *construct* the representations themselves and be given opportunities to *translate* between different representations. The point about learners needing repeated practice in translating between different representations and to the real world was also underscored by McDermott [8].

Table 1: Typical visual representations associated with specific physics topics.

| Physics Topics | Visual Representations |
|---|---|
| Kinematics | Motion diagrams |
| Forces & Dynamics | Free-body diagrams |
| Energy | Energy bar charts |
| Field | Field line diagrams |
| Electric circuits | Electrical circuit diagrams |
| Geometrical optics | Ray diagrams |
| Waves | Wavefront diagrams |
| Quantum Physics | Energy level diagrams |

In the next section, we describe how we used MRs to help student consolidate their understanding of physics concepts in an A-level mechanics revision lesson. The context of the problem involved the motion of a ball thrown upwards in air and studying how the associated physical quantities changed along its flight.

**The Multiple Representations Revision Lesson**

*Overview of Lesson*

The lesson began with the teacher throwing an actual tennis ball in the air and asking the class to think about how the different associated physical quantities (e.g. displacement, velocity, acceleration, force, momentum, energy) changed along the ball's flight on its way up and down. As part of the preliminary analysis, the teacher instructed students to ignore the effects of air resistance, and pay particular attention to the following instances of time: just after release from the hand ($t = 0$ s); on its way up ($t = 1$ s and $t = 2$ s); at the highest point ($t = 3$ s); on its way down ($t = 4$ s and $t = 5$ s) and just before landing on hand ($t = 6$ s). Different groups of students were then assigned to look at different aspects of the ball's motion, using a range of representations: vector diagrams, free-body diagrams, verbal description, equations and graphs (see the Appendix for the handout of the lesson). For the graphical representation, students were asked to draw two sets of graphs: the first set involving graphs of displacement, velocity and acceleration against time/displacement and the second involving graphs of energy against time/displacement. Figure 1 presents the interconnections among the different representations that describe the phenomenon of the ball's motion.

Every student was fully engaged in the discussion. The teacher, acting as a facilitator, asked guiding questions to help clarify, probe and extend students' thinking. The handout for the lesson served to give a broad overview of the lesson while defining clearly the specific task that each group had to undertake. After the small group discussion, a student from each group presented their work to the rest of the class, with the teacher helping the students to draw connections between the different perspectives and ensuring consistency in the way the concepts are understood using the different representations. The teacher also guided students to synthesize and consolidate their learning of the related concepts (e.g. Newton's laws of motion, conservation of energy).

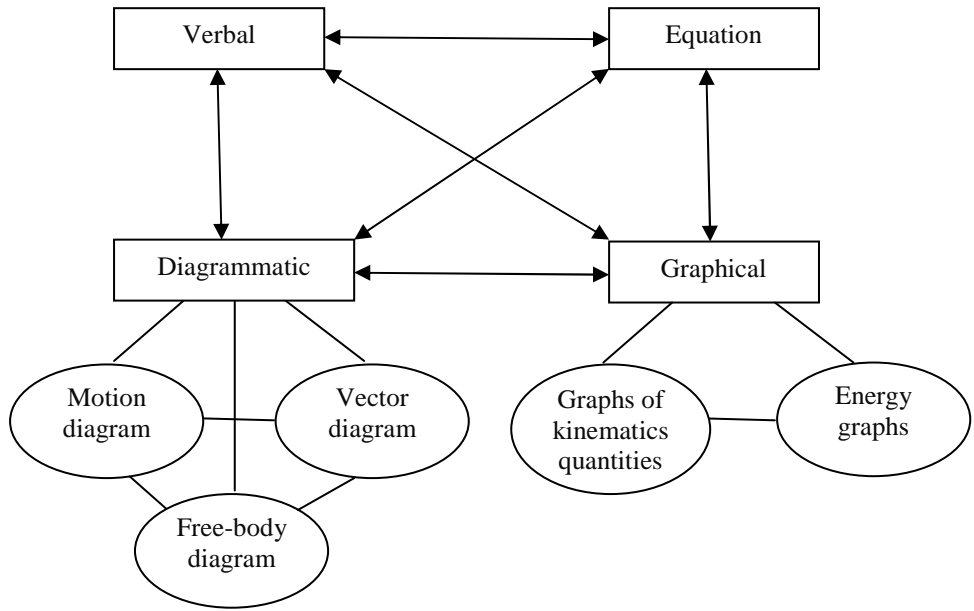

Figure 1: The interconnections among the different representations that describe the phenomenon.

*Diagrammatic and Verbal Representations*

The discussion began with students presenting the diagrammatic representation since it is most concrete and intuitively associated with the actual motion of the ball. Subsequently, students can use this to anchor their understanding of the other representations. While students may depict the motion of the ball as a two-dimensional representation on paper, as seen from Figure 2(a), it would be important to highlight that the motion of the ball was actually one-dimensional. Two separate motion diagrams (one for the motion on the way up and the other on the way down) could therefore have been drawn, as shown in Figure 2(b). They would be identical and serve to highlight the symmetry of the upward and downward motions. On the way up, one would expect the balls to be drawn at successively closer distances apart for every 1-second time interval, since the ball slows down due to Earth's gravitational pull. To scaffold the construction of the *s-t*, *v-t* and *a-t* graphs, a table showing the relative values of the kinematics quantities at various times could also be drawn, as shown in Figure 2(b).

To check for student understanding, it would be instructive to have them consider the position the ball would reach at half the time it takes to reach the maximum height. That is, at $t = 1.5$ s, would the ball be at a point *lower*, *higher* or *exactly at* the mid-way mark of the ball's motion? Student would need to be able to explain that this point would be actually greater than the mid-point position due to the slowing down the balls' motion on its way up, in which it covers more distance in the first 1.5s.

Besides the motion diagram (i.e. trajectory of the ball's position at successive equal intervals of time), we also required the students to include vector quantities at each of the ball's positions. One group of students worked on the *kinematics* perspective by drawing velocity and acceleration vectors (vector diagram) and another group considered the *dynamics* perspective by drawing the forces acting on the ball (free-body diagram). Students in the latter group were not able to rationalise why an upward moving ball had no 'force' vector pointing upward. Here, it may be helpful to highlight to the students that while that was no upward 'force' vector, we can in fact draw 'momentum' vectors pointing upwards, with decreasing magnitude. Hence this explains why the force causing the change in momentum must be pointing in the opposite direction since the force is directly proportional to the rate of change of momentum.

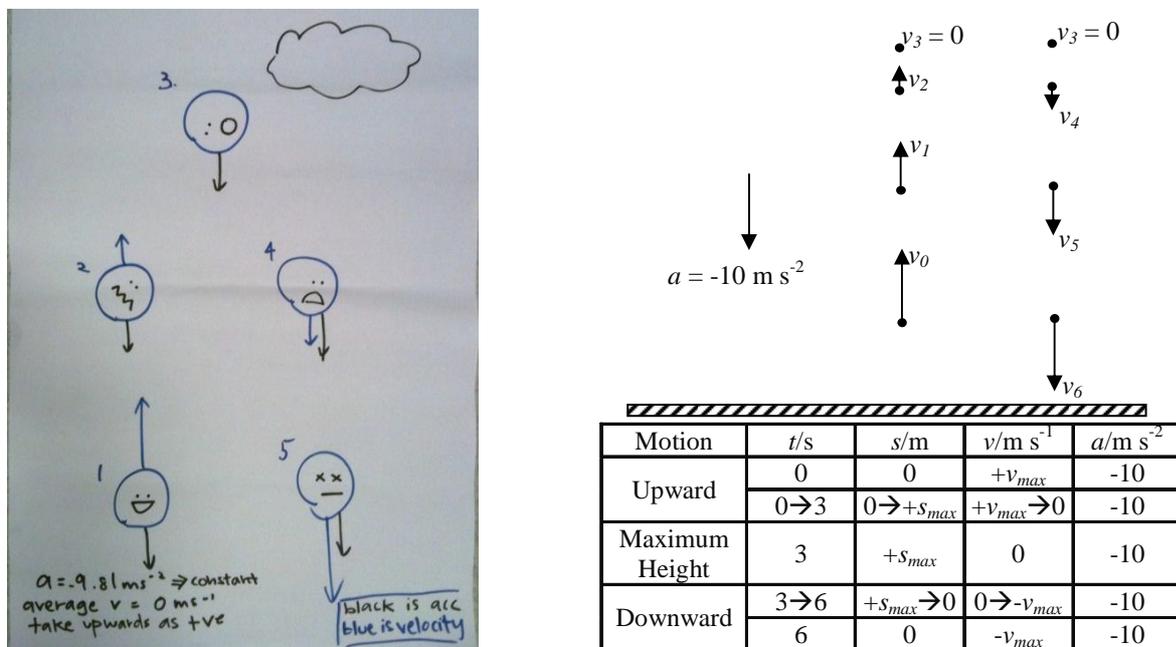

Figure 2: (a) Students' depiction of the motion of the ball using a diagrammatic representation. (b) Suggested diagrammatic and tabular representations to depict the ball's motion.

To sharpen students' thinking and the use of language, we specifically looked out for consistency in the verbal and diagrammatic representations. For example, we would expect students to describe the ball's motion as: *the ball slows down as it travels upwards at a constant rate (of nearly 10 ms$^{-2}$), stops momentarily at the highest point, and speeds up at the same constant rate on its way down.* In explaining the cause of the ball's motion, we would expect students to refer to the force of gravitational pull on it by the Earth. Since we have neglected air resistance, $g$ can be considered essentially constant near the Earth's surface.

*Graphical and Mathematical Representations*

Student difficulties in connecting graphs to physical concepts and to the real world – such as constructing graphs which resemble the spatial appearance of the motion – for the topic of kinematics have been well documented [9]. When discussing the construction of graphs for the kinematics quantities (i.e. *s-t*, *v-t* and *a-t* graphs), it would be useful to have students explicitly articulate their chosen conventions. For example, for the graphs drawn by the students (see Figure 3), they had chosen upward motion as positive and the reference point, $s = 0$ m, at the starting point of the ball's motion. The significance of the mathematical relationships between the graphs (i.e. $v = ds/dt$ and $a = dv/dt$) need to be emphasised so that students are able to relate how one graph is obtained from the other. For example, students should be able to see that, on the *s-t* graph, (a) positive slope implies motion in the positive direction, (b) zero slope implies a state of rest, and (c) negative slope implies motion in the negative direction.

While trying to explain why the velocity at the mid-point was closer to the initial value, a student gestured that the ball rose to smaller and smaller increments in height and hence thought that *"the velocity was decreasing at a decreasing rate"*. This error in thinking persisted despite the teacher directing his attention to the *v-t* graph being linear. The student obviously was confused over the decrease in velocity with the *rate* of decrease. This underscores the need to teach students to discriminate between the slope and height of a graph. On an *s-t* graph, students should be able to relate that: (a) straight lines imply constant velocity, (b) curved lines imply acceleration, and (c) an object undergoing constant acceleration traces a portion of a parabola. The skill of being able to verbally describe the shape of graphs is also important (e.g. linear graph, upward-opening parabola, sideway-opening parabola, etc).

During the lesson, students wrote down the general equations of motion. It would be instructive for the teacher to ask if the situation could be characterised more specifically, with numerical values for the set of equations. Students could be guided to obtain the initial velocity by following this simple line of argument: since the ball takes 3 s to reach zero velocity, the initial velocity must have been 30 m s$^{-1}$ since the speed decreases by 10 m s$^{-1}$ each second. Hence, the equations of motion become: (a) $s = 30t – 5t^2$ (b) $v = 30 – 10t$ and (c) $v^2 = 30^2 – 20s$. For each equation, a corresponding graph can be drawn and specific features of each graph that relates to the ball's trajectory can be highlighted, as shown in Figure 3(b). Indeed we need to give students ample opportunities to practise selecting the appropriate feature of a graph to obtain the information required. For example, we could ask the following guiding questions to lead students to realise how the same information can be obtained using different graphs by focussing on different features from each graph: (a) What is the gradient at the highest point on the s-t graph? (b) What is the velocity reading on the v-t graph at that time? (c) Is it possible to obtain the velocity at the highest point using the a-t graph?

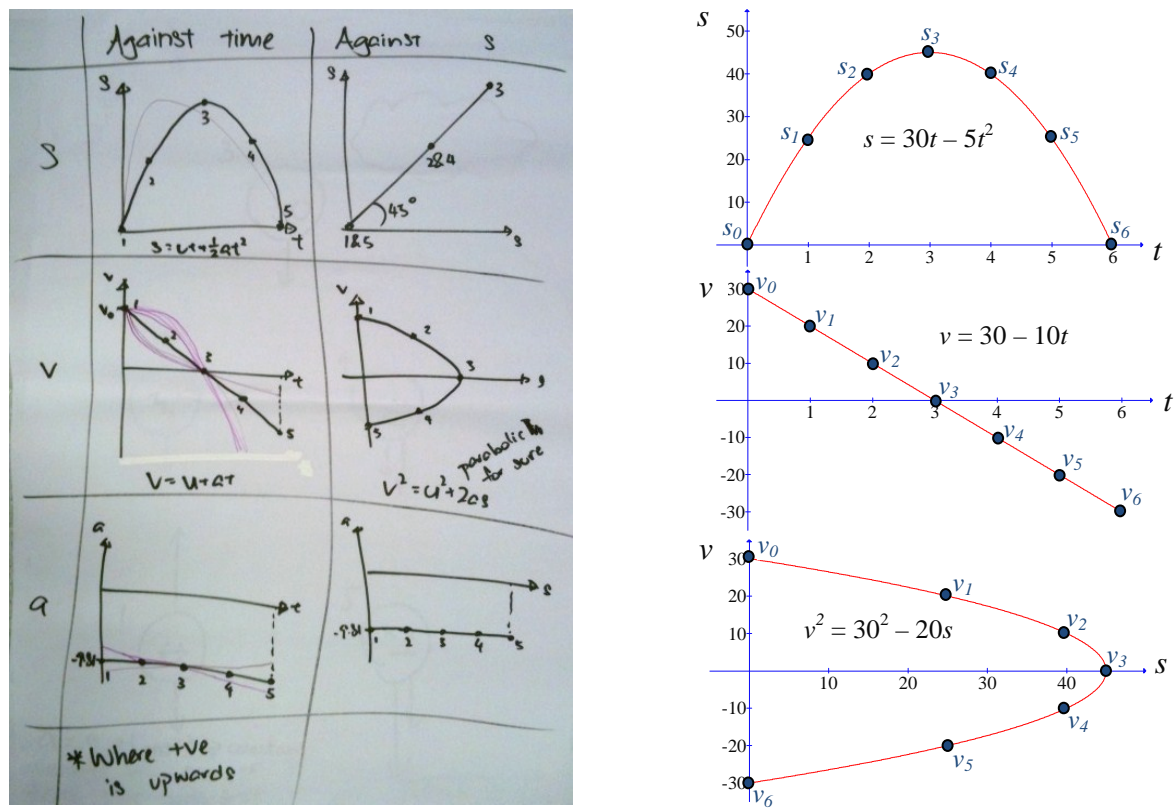

Figure 3: (a) Students' depiction of the motion of the ball with the graphical representation. (b) The graphs depicting the equations of motion for this specific situation.

When dealing with the *v-s* graph, students may be confused over whether the velocity at the mid-position (i.e. at *s* = 22.5 m) on the *s-t* graph is that of the velocity at the mid-time (i.e. at *t* = 1.5 s) on the *v-t* graph. To address the confusion, we need to stress to students that the mid-time does not coincide with the mid-position due to the accelerated motion, as previously discussed. In fact, while the velocity at mid-time is simply $v = 15$ m s$^{-1}$, the velocity at mid-position corresponds to the vertical line on the *v-t* graph that bisects the positive or top half of the area into two equal parts. Earlier we discussed that the position of the ball at mid-time was greater than the mid-point position, as can be seen clearly from the *s-t* graph. Alternatively, we could also direct students' attention to see that consistent with this fact is the observation that the left portion of the area under the *v-t* graph at mid-time is larger compared to the right portion.

It would be instructive to have students reason out why the *KE-s* and *PE-s* graphs are linear, in contrast with the *KE-s* and *PE-s* graphs which are curves. Students are familiar with the equation for *PE* near the Earth's surface given as *mgh*. This relation can be used to explain why the *PE-s* graph is a straight line with positive slope passing through the origin. As for the *KE-s* graph, there are two ways to look at it. First, from the equation of motion: $v^2$ is linearly related to *s*. We can therefore infer that *KE* is linearly related to *s* since the mass of the ball remains constant. The other way is by conservation of energy: since the total energy is constant, *KE-s* graph will have to be linear too so that at any position, the sum of KE and PE is constant.

*Ball's Motion with Air Resistance*

For the discussion of the ball's motion with air resistance, an interesting point to consider is the top of the trajectory: this a turning point and because it is momentarily at rest, the air resistance will have no effect at this point. The following are some probing questions we could ask students: (a) What effect does air resistance have on the resultant acceleration of the ball? (b) At which point(s) of the ball's motion is $a = 10$ m s$^{-2}$? (c) How would the graphs of the kinematics quantities be affected by air resistance? and (d) How does the time taken to travel up compare with that to travel down?

On the problem of whether it takes the ball longer to travel up or down when air resistance is considered, students intuitively argue that since the net acceleration of the ball on its way up is greater than that on its way down, it must take a shorter time to go up. This explanation is flawed because the initial speed of the ball upward is different from the final speed of the ball downwards, so that it is difficult to say whether it takes longer to decrease the initial speed to zero or to increase the speed to the final value (assuming terminal velocity is not reached). A better way would be to look at the average velocity on its way up and down. Knowing that the air resistance is acting on the moving ball, the *KE* of the ball would decrease with time and so it is reasonable to infer that the average velocity upward is greater than the average velocity downward. As average velocity is the change in displacement over time, and the change in displacement is same magnitude going up and coming down, then the time to go up must be smaller.

A student, drawing on her knowledge that average speed is the total distance travelled divided by the duration of travel, mistakenly suggested that the average velocity is the *total* displacement divided by the duration. This led to an interesting discussion on the meaning of total displacement. If we cut the duration into finer intervals, would the total displacement then change as the total displacement would have to be the vector sums of all the individual displacements at each interval!

*Feedback from Student*

The feedback about the lesson, summarised in Table 2, indicates that the majority of students either strongly agreed or agreed to the benefits of the lesson. Students generally found the lesson interesting and interactive, and requested for such lessons to be conducted for other topics as well. In particular, they felt that the lesson helped clarify and strengthen their understanding of the related concepts and that they were now better able to apply these concepts in answering questions. Indeed, the lesson also helped some to stretch their learning and they were satisfied with the fact that they were able to reason out things for themselves: for example, how to reason out why time taken to go down is greater than time taken to go up when air resistance is considered. Some of the comments were:

- *Really good way to learn and if able to allocate time appropriately, would recommend strongly for future batches to learn this way.*
- *I managed to derive the v-t graph with air resistance while discussing with peers! I never knew how the graph looked like before.*

Table 2: Survey results for lesson which involved learning with multiple representations (N = 25).

| | Question | Strongly Agree | Agree | Disagree | Strongly Disagree |
|---|---|---|---|---|---|
| 1. | I would recommend learning Physics using this strategy. | 36% | 56% | 8% | - |
| 2. | I gain a deeper understanding of the mechanics concepts. | 44% | 56% | - | - |
| 3. | It helps me realize some of my inappropriate understanding. | 24% | 76% | - | - |
| 4. | It helps me appreciate mechanics using different representations. | 20% | 76% | 4% | - |
| 5. | I discover something new in this session. | 36% | 60% | 4% | - |

**Supporting and Extending Learning with Simulations**

Simulations and multimedia learning environments [10-11] offer the unique advantage of combining different representations in one interface. In particular, dynamic presentations offer the possibility to connect representations not only by integrating them, but also by linking them so that a change in one representation is concurrent with a change in another representation. This helps learners establish relationships between the representations. Besides supporting learning in the classroom, simulations can also help consolidate and extend student learning beyond the classroom and allow motivated students to test out the model of the physical phenomena in different scenarios at their own time.

We have created a simulation[2] using *Easy Java Simulations* (Ejs) [12] that can be used to support active learning with MRs for the lesson. The simulation is able to dynamically plot out the shapes of different graphs with the diagrammatic view of the motion of the ball. One of the design considerations for the simulation is that it allows only selected variables to be visible so that it focuses learning without unnecessary cognitive overloading. For example, if we want to focus students' attention on how the energy (TE, KE & PE) changes it would be useful to first introduce energy bar charts, before introducing students to the actual plots of the energy graphs (see Figure 4). Notice also that the simulation programme allows the trace of the ball to be drawn, akin to the motion diagram discussed earlier.

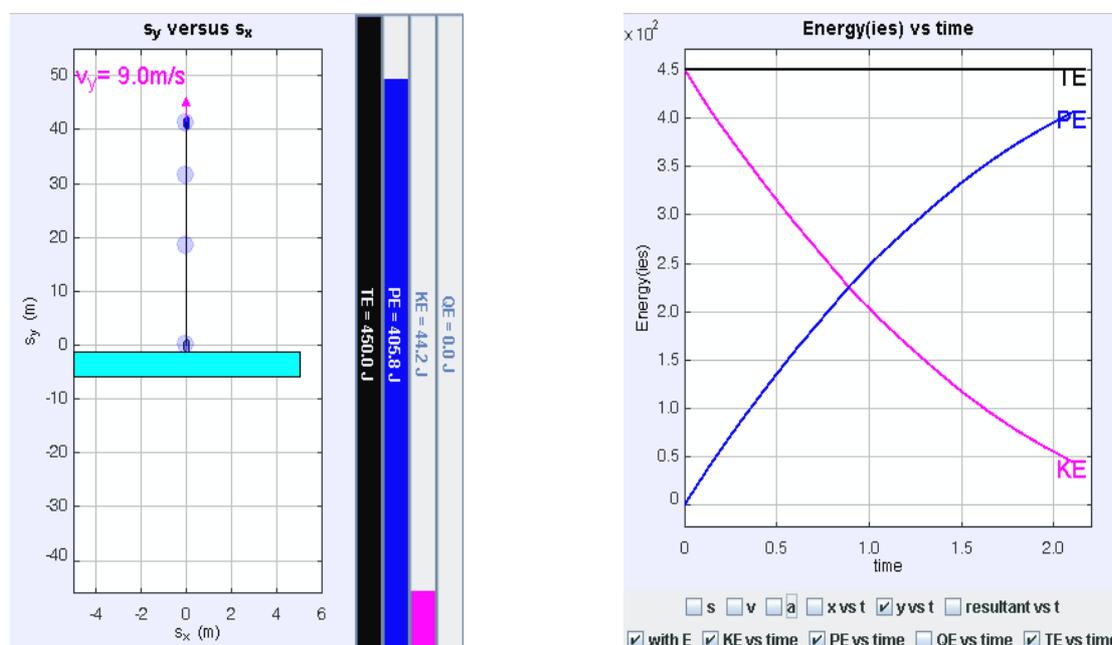

Figure 4: A screen capture of the simulation programme showing the diagrammatic and graphical representations of the ball's motion.

As an extension of the classroom lesson, we could have students explore the motion of the ball for more than one bounce and at different conditions of elasticity of impact with the ground. For example, the simulation programme allows for easy comparison of the graphical plots of various quantities when air resistance is present to the situation without air resistance (see Figure 5). Students can be guided to predict what the graphs with air resistance might look like, and checking with those obtained from the simulation.

---

[2] The simulation can be assessed and downloaded at the NTNU Virtual Physics Laboratory Discussion Forum at the following website: http://www.phy.ntnu.edu.tw/ntnujava/index.php?topic=1940.0.

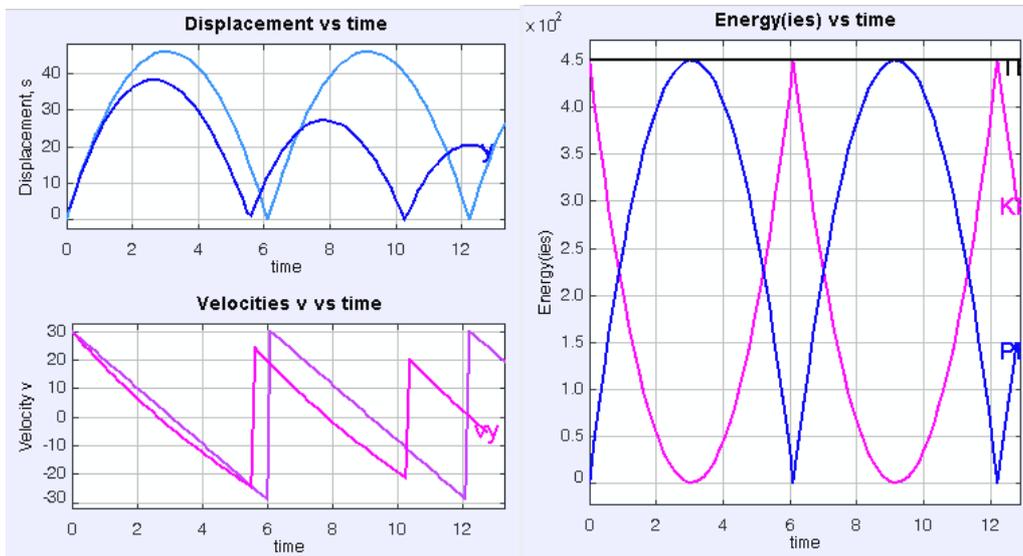
Figure 5: Comparison of graphs for the bounce of a ball with and without air resistance.

**Conclusion**

We have described how learning with multiple representations can be used within and beyond the classroom. We believe that providing students the opportunity to construct multiple representations of physical phenomena and deliberating over the connections between these representations can help students develop a deeper and more coherent understanding of physics concepts and better tackle physics problems.

<div style="text-align: right">**Appendix**</div>

**Handout for Multiple Representations Revision Lesson**

**Instructions:**
A ball is thrown vertically upwards at time t = 0s. It is caught at the height of release on the way down at time t = 6s. Do the following exercises, paying attention particular attention to the following moments:
1.  just after release from the hand (t = 0s);
2.  on its way up (t = 1s  &  t = 2s);
3.  at the highest point (t = 3s);
4.  on its way down (t = 4s  &  t = 5s);
5.  just before landing on hand (t = 6s).

**A       Vector Diagrams**
Draw vector diagrams to show the position of the ball at equal time intervals with its corresponding velocity vector (blue) and acceleration vector (black).
Think about how you would describe the velocity and acceleration of the ball on its way up, at the top and on its way down.

**B       Force Diagrams**
Draw free body diagrams showing momentum and all forces (use different colour for different forces, use red for net force) acting on the ball for the upward and downward motion.
Think about how you would describe the change in momentum of the ball on its way up, at the top and on its way down.

**C       Graphs**
Draw graphs of displacement, velocity, acceleration with time/displacement to show the motion of the ball.

**D       Energy graphs**
Draw the KE (red) and PE (blue) graphs wrt time and displacement for the motion of the ball.
Think about how you would describe the transformation of energy of the ball in its flight.

**E       Equations**
Write down the equations of motion that describe the motion of the ball.
What is the average velocity of the ball?
What is the average acceleration of the ball?
With air resistance, does it take longer to go up or come down?

10